\documentstyle[aps,preprint]{revtex}
\input{epsf}
\begin{document}

\preprint{preprint from Electron.Lett.{\bf 7}, 27.3.97}

\title{Interferometry with Faraday mirrors for quantum cryptography}

\author{H.Zbinden, J.D.Gautier, N.Gisin, B.Huttner, A.Muller, and W.Tittel}

\address{University of Geneva, 
         Group of Applied Physics, 
         20,Rue de l'Ecole de Med\'ecine, 20, 
         CH-1211 Geneva 4, Switzerland\\
         Tel: +41 22 702 6883; Fax: +41 22 781 0980\\
         email: hugo.zbinden@physics.unige.ch}

\date{\today}
\maketitle


\begin{abstract}
Quantum Cryptography over 23km of installed Telecom fiber using a novel 
interferometer with Faraday mirrors is presented. The interferometer needs no 
alignment nor polarization control and features 99.8\% fringe visibility. A secret 
key of 20kbit length with a error rate of 1.35\% for 0.1 photon per pulse was 
produced. 
\end{abstract}


\newpage

In cryptography, safety can be obtained by exchanging a secret key between the two 
users, Alice and Bob. In quantum cryptography (QC) the key is exchanged through a 
quantum channel. Its security is based on the fact that any measurement of a quantum 
system will inevitably modify the state of this system. Therefore an eavesdropper, 
Eve, might get information out of a quantum channel by performing a measurement, 
but the legitimate users will detect her and hence not use the key. In practice the 
quantum system is a single photon propagating through an optical fiber, and the key 
can be encoded by its polarization or by its phase, as first proposed by Bennett and 
Brassard \cite{one}. In 1992 quantum cryptography was for the first time 
experimentally 
demonstrated over 30 cm in air with polarized photons \cite{two}. Since then, several groups 
presented realizations of both, the polarization \cite{three} and the phase coding scheme in 
optical fibers over lengths of up to 30 km \cite{four,five}. 

However, all quantum cryptography systems face two main difficulties. The first 
problem is the need of continuous alignment of the system. In polarization-based 
systems, the polarization have to be maintained stable over tens of kilometers, in order 
to keep aligned the polarizers at Alice's and Bob's. In interferometric systems, usually 
based on two unbalanced Mach-Zehnder interferometers, one interferometer has to be 
adjusted to the other every few seconds to compensate thermal drifts \cite{four}. 
The second problem is the high noise of photon counters at 1300nm which essentially determines 
the error rate (ER) in the key.

In this letter, we present the creation of a secret key over 23km of installed Telecom 
fiber using a recently introduced interferometric system with Faraday mirrors \cite{six}. 
This phase-coding setup needs no alignment of the interferometer nor polarization control. 
It features excellent fringe visibility and stability. Moreover we show that the 
performance of Ge-APD photon counters can be considerably improved using a fast 
active biasing electronic.  

Our novel quantum cryptography scheme is shown in Fig.1. In principle we have an 
unbalanced Michelson interferometer at Bob's (beamsplitter C2) with one long arm 
going to Alice. The laser pulse impinging on C2 is split in two pulses P1 and P2. P2 
propagates through the short arm first (mirror M2 then M1) and then travels to Alice 
and back, whereas P1 is going to Alice first and passes through the short arm on its 
way back. Both pulses run exactly the same path and will interfere at C2. To encode 
their bits , Alice is acting with her phase modulator (PM) only on P2 (phase shift  
$\phi_a$), whereas Bob lets pulse P2 pass unaltered and modulates the phase of P1 (phase 
shift $\phi_b$). If no phase shifts are applied or if the difference 
$\phi_a-\phi_b=0$, then the interference 
will be constructive. The interference will be destructive when $\phi_a-\phi_b=\pi$ and no 
light will be detected by detector $\mbox{D}_0$. Since the interfering pulses travel the same 
path, the interferometer is automatically aligned. Thanks to Faraday mirrors, a 
$45^\circ$ Faraday rotator (FR) in front of an ordinary mirror, any birefringence in the 
interferometer is 
compensated and no polarization control is necessary \cite{seven}. We measure the ratio of the 
count rates for constructive and destructive interference that correspond directly to 
the contribution of imperfect interference to the ER in the key, which we denote by 
$\mbox{ER}_{opt}$. 
When we apply a phaseshift at Bob's piezo-optical modulator we obtain an 
attenuation of 30db$\pm$1db, while when we apply the phaseshift at Alice's 
integrated phase shifter the extinction is 27db$\pm$1db. This values were 
reproducible within the 
given errors over weeks. An extinction of 30db correspond to a classical fringe 
visibility $\rm V = (I_{max} - I_{min}) / (I_{max} + I_{min})$ of 99.8\%. The measured 
values 30db and 27 db result in 
$\mbox{ER}_{opt}$ of 0.1\% and 0.2\%, respectively. The average, decisive 
for the key creation is therefore 0.15\%. Replacing any Faraday mirror by an ordinary 
mirror, the extinction is strongly fluctuating and can be reduced to 20db. If two Faraday 
mirrors are removed, essentially no interference is visible. 

We work with the BB92 protocol \cite{eight} using two states. Alice and Bob choose at 
random 0 or $\pi$ phase shifts, defined as bit values 0 and 1. If a detection 
i.e. constructive interference occurred, Alice and Bob know that they applied the same 
phase shift, and that they had the same bit value. Pulses where no detection occurred 
are simply disregarded. To prevent Eve to split the pulses to get information on the 
applied phase, they must contain 1 photon at most. Therefore strongly attenuated 
laser pulses with an average number of photons $\mu$ well below 1 are used. The pulses 
leaving Bob do not yet carry a phase information. The information is in the phase 
difference of two pulses P1 and P2 leaving Alice. The attenuator (A) is set in a 
manner that the weaker pulse P2, that passed already Bob's delay line, has 0.05 
photons in average. This is equivalent to an intensity after interference of $\mu=0.1$ for 
the pulse pair. Eavesdropping on 2-states system in general \cite{eight} and our setup in 
particular \cite{nine} is out of the scope of this article. Our setup could be quite easily 
adapted to a four states protocol BB84 \cite{one}. We tested that using also 
$\pi/2$ and $3\pi/2$ phase shifts the same excellent performances of the interferometer are 
obtained.

Our photon counter is a $\mbox{LN}_2$-cooled Ge-APD (NEC NDL5131) used in the 
Geiger 
mode. The bias voltage of diode is the sum of a DC part well below threshold and a 
2ns long almost rectangular pulse of 7.5 V amplitude that pushes the diode 1.4 V over 
threshold, when the photon is expected. This increases considerably the quantum 
efficiency without excessively enlarging the noise and reduces the time jitter to below 
100 ps. In a time window of 300 ps we get 22 and 7 darkcounts per 1 million pulses 
for efficiencies of 20\% and 10\%, respectively. This will lead to a detector 
induced ER ($\mbox{ER}_{det}$) of 0.72$\pm$0.13\% at 10\% efficiency ($\mu=0.1$, 10 db 
loss in the line.). 
 
The heart of our experiment is a SRS 535 delay generator (SRS) at Bob's. It beats at 
1KHz and triggers the laser, Bob's phase modulator (PM), the active biased photon 
counter ($\mbox{D}_0$), and Bob's computer. The 1300nm Fujitsu-DFB-laser driven by an 
Avtech pulser delivers 300ps pulses (FWHM). The phase modulator is a fiber 
wrapped around a piezoelectric-tube. It is driven by a sinus function from a SRS DG 345 
function generator (FG). The modulation frequency of the piezo of about 10kHz 
is high enough since the time delay between the outgoing and incoming pulses is 
about 230 $\mu$s. 
Only if the computer gives a logical 1 to the and-gate at the external 
trigger input of the function generator a phase is applied. The pulses run down a 22.8 
km long Telecom link between Geneva and Nyon, Switzerland, featuring 8.6 db loss. 
On Alice's side, the pulse P1 detected by DA (Newport AD-300/AC) triggers the 
phase modulator driven by another SRS function generator and the computer. In 
order to switch within the 250ns delay between the two pulses, a Thomson 4GHz 
LiNbO3 waveguide phase modulator is used. 

Back at Bob's, the interfering photon directly runs to the detector 
$\mbox{D}_0$ via the 10 db 
coupler C1 to limit the losses. The photon counter electronics is precisely triggered to 
ensure that the arrival of the photon and the biasing of diode coincide within 100 ps. 
Every detection is registered by Bob and assigned to the number of the pulse after the 
beginning of the measurement. Alice and Bob had 100 files of 65535 bit of random 
numbers. This numbers have been generated by an apparatus based on thermal noise 
of an electrical resistor \cite{ten}. After the measurement the results are compared with the 
random lists of Alice and Bob in order to determine the ER. The results are 
summarized in Table 1.
\vspace{0.5cm}

\tabcolsep2mm
\begin{tabular}{|c|c|c|c|c|c|}
\hline
Average photon & measured ER & $\mbox{ER}_{det}$ & $\mbox{ER}_{opt}$ & Length 
of key & Bit rate \\
number $\mu$ per & (\%) &  (\%) &  (\%) & (bit) & (Hz)\\
pulse pair & & & & &\\
\hline
0.2 & $0.5\pm0.1$ & $0.4\pm0.07$ & $0.15\pm0.03$ & 2980 & 0.9\\
\hline
0.1 & $1.35\pm0.08$ & $0.81\pm0.14$ & $0.15\pm0.03$ & 20142 & 0.5\\
\hline
\end{tabular}  

\vspace{0.5cm}
To our knowledge, the achieved ER's are the lowest ever obtained over a distance of 
more than 20km. The measurement with $\mu$ =0.1 lasted more than 11 hour and no 
realignment was performed. The obtained ER is slightly higher than the sum of 
$\mbox{ER}_{det}$ and $\mbox{ER}_{opt}$. 
We believe that this is rather due to variations in the photon counter, its 
electronics and timing, than in any fluctuations of the interferometer. We tested 
another setup where the photon counter was triggered by the laser pulse at the trigger 
output (T) running down another fiber to Alice and back. This measure can reduce the 
susceptibility to temperature variations, and therefore improve the stability under 
difficult environmental conditions. The achieved bit rates are quite low. This is simply 
due to the low pulse rate, that could be increased by replacing the piezoelectric 
modulator and adapting the computer steering. Please note, that the timing of Alice's 
apparatus can be pre-adjusted in the lab, and will not change, even if the apparatus is 
plugged to another fiber to communicate with a third party. The timing of the Bob's 
apparatus, especially of his photon counter has to be adjusted once for every link. 
Synchronization between Alice and Bob is automatically assured by the pulse P1. 

In conclusion we have shown that the introduced interferometric QC-setup using 
Faraday mirrors features impressive stability and a fringe visibility of 99.8\% 
without any alignment. We produced a secret key of 20 kbit length with a ER of 1.35\% for 
0.1 photon per pulse. This low ER is also due to an actively biased germanium photon 
counter with reduced noise, compared to a passive system. Equipped with faster 
components and electronics to increase the bit rate and Peltier cooled InGaAs-
detectors, such a system has the potential to make QC practical for Telecom 
applications.  

We would like to thank the Swiss Telecom for financial support and for placing at our 
disposal the Nyon-Geneva optical fiber link. We appreciate the stimulating discussions 
with our colleagues within the TMR network on the Physics of Quantum Information.


\newpage

\begin{figure}[p]
\hspace*{1cm}
\epsfxsize=14cm \epsfbox{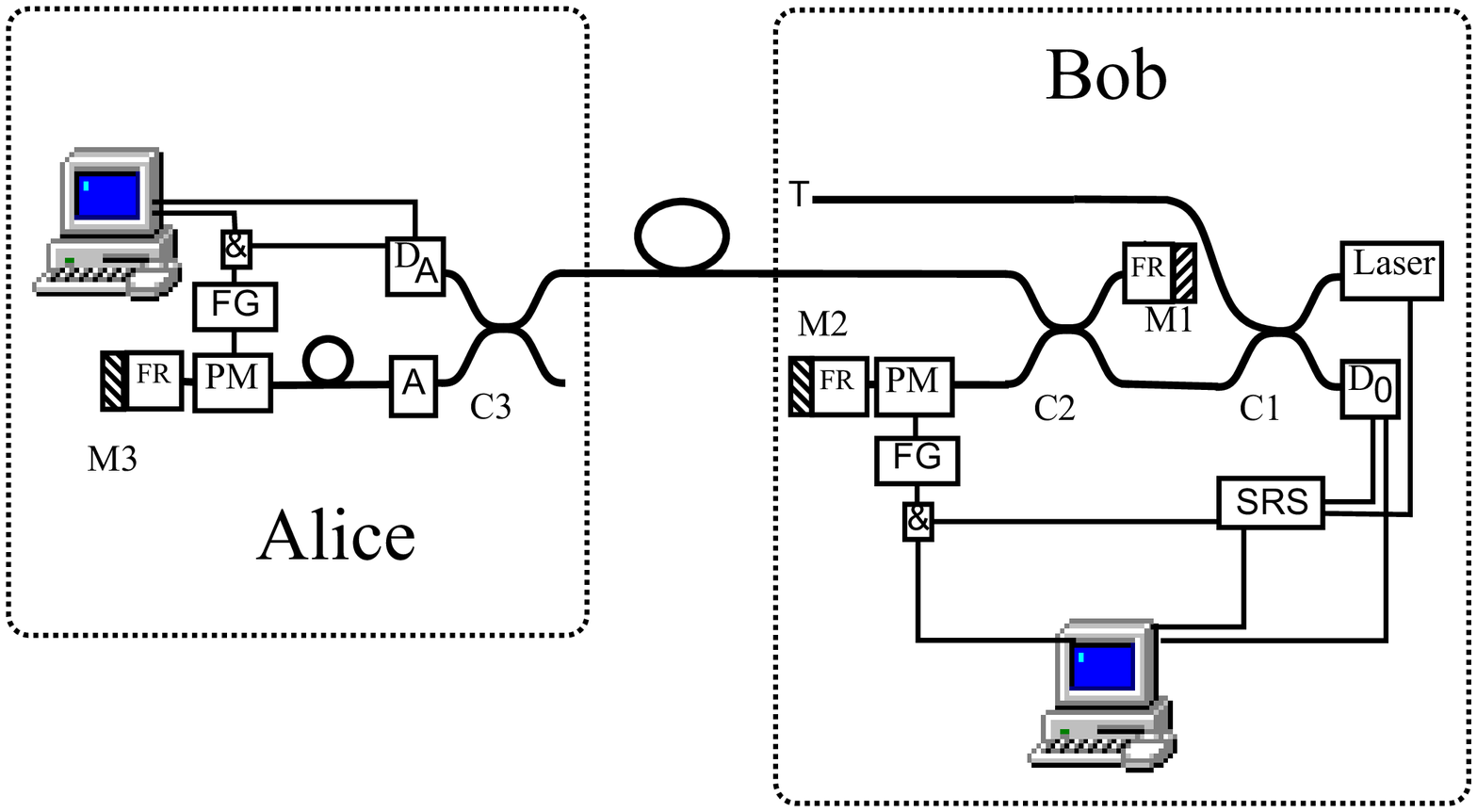}
\caption{Schematic diagram of the interferometric QC system with
Faraday mirrors.}
\end{figure}

\end{document}